\documentclass[onecolumn]{emulateapj} 
\usepackage{amsmath}
\usepackage{multirow}
\usepackage{epsf}
\usepackage{rotating}
\usepackage{lscape}
\usepackage{longtable}
\usepackage{flushrt}

\usepackage{graphicx}
\usepackage{graphpap}

\def\dm{$^{\prime}$}
\def\d{$^{\circ}$}

\def\etal{{et al.}}
\def\chandra{{\em CHANDRA}}
\def\xmm{{\em XMM-Newton}}
\def\asca{{\em ASCA}}

\lefthead{Donnelly \etal}
\righthead{Flux Limited Atlas of Galaxy Cluster Temperature Maps}

\slugcomment{submitted to {\em The Astrophysical Journal} July 23, 2003}
\begin{document}

\title{A Flux Limited Atlas of Galaxy Cluster Temperature Maps}

\author{R.~Hank~Donnelly, C.~Jones, W.~Forman, E.~Churazov, M.~Gilfanov}

\affil {Harvard-Smithsonian Center for Astrophysics, 60 Garden Street,
Cambridge, MA 02138, USA.\\ e-mail: rdonnelly@cfa.harvard.edu} 

\centerline{\date}
\begin{abstract}
An atlas of gas temperature maps is presented for a flux limited
catalog of galaxy clusters. The sample of clusters is based on the
Edge \etal\ (1990) sample, with the inclusion of five additional
clusters, all with fluxes $f_X(2-10$ keV$)\ge 1.7\times 10^{-11}$ ergs
sec$^{-1}$ cm$^{-2}$, drawn from a variety of other sources. The
temperature maps are derived from \asca\ GIS observations using a
common methodology to correct for the Point-Spread Function and
calculate the local projected gas temperature in such a way so as to
make each cluster directly comparable to all others in the sample.
Variations in the temperature distribution, when present at $> 90$\%
confidence, are characterized by their severity and extent.  We find
that 70\% of the clusters in our sample have significant variations in
the projected gas temperature. The presence of these variations
increases with increasing luminosity, as does the spatial scope and
severity within a cluster.  For a more limited sample we find that one
third of clusters with temperature structure have radio halos.  The
high rate of occurance of structure emphasizes the need for caution
when using clusters to measure cosmological parameters.
\end{abstract}


\section{INTRODUCTION}
The hot intracluster medium (ICM) is the dominant component of the
luminous baryons in galaxy clusters. This diffuse gas, typically
heated to tens of millions of degrees Kelvin, comprises about 25\% of
the total mass and approximately five times the mass in the galaxies
(Blumenthal \etal\ 1984; David \etal\ 1990; Arnaud \etal\ 1992; White
\etal\ 1993; David, Jones \& Forman 1995; White \& Fabian 1995; Allen
\etal\ 2003). Multiple studies of the X-ray surface brightness of the
ICM (Abramopoulous \& Ku 1983; Jones \& Forman 1984; Fabricant \etal\
1986; Edge \& Stewart 1991a, 1991b; Mohr, Fabricant \& Geller 1993;
Slezak, Durret \& Gerbal 1994; Durret \etal\ 1994; Buote \& Xu 1997;
Jones \& Forman 1999) have shown that a substantial percentage (40\%)
of clusters are dynamically active and have observationally confirmed
the suggestion that ``the present is the epoch of cluster formation'' (Gunn
\& Gott 1972).

However, characterizing the dynamical state of a cluster by this
approach is limited somewhat by the viewing geometry and the
relatively rapid return of the surface brightness to a relaxed
distribution (Schindler \& M\"uller 1993; Ricker 1998). In contrast,
variations in the projected gas temperature due to merger events can
persist for longer timescales and are detectable even when the merger
is along the line of sight. Thus with the advent of spatially resolved
spectroscopic instruments (e.g. \asca, and more recently \chandra\ and
\xmm), we are more sensitive to detecting merger events and are able
to test the nominal assumption of isothermality (see e.g. Markevitch
\etal\ 1998). Deviations from this assumption have important
consequences for determinations of cluster masses, and thence to the
determination of the fractional amount of matter contained in
gravitationally bound systems as well as for testing cosmological
models (Reiprich \& B\"ohringer 2002).

In this paper we present an atlas of the gas temperature distributions
for 58 galaxy clusters comprising a flux limited sample.  By analyzing
this sample using a common set of techniques and levels of confidence
we can make direct comparisons between the various members and explore
the presence of substructure at the current epoch. Our sample also
highlights potential complications for higher redshift objects and can
be used as a comparison sample for simulations testing cosmological
parameters.

Section 2 describes the development of our sample;
Section 3 outlines the issues involved in the
derivation of robust gas temperature maps from \asca\ observations,
and describes in general the methodology applied.
Section 4 details the global results for each cluster
and presents the gas temperature maps for each cluster sorted by
decreasing luminosity.  Section 5 describes correlations
of the temperature structure with luminosity and the presence of radio
halos. We have applied the WMAP cosmology (Bennett \etal\ 2003)--
$\Omega_m=0.27$, $\Omega_\Lambda=0.73$ and $H_0=71$ km sec$^{-1}$
Mpc$^{-1}$-- to derive all distance related quantities.

\section{THE SAMPLE\label{sec:sample}}
Our sample uses the flux limited catalog of Edge \etal\ (1990; revised
in 1992) as its foundation. These 55 clusters from the {\em EINSTEIN},
{\em HEAO-1}, and {\em EXOSAT} data sets have 2-10 keV fluxes greater
than $1.70\times 10^{-11}$ ergs cm$^{-2}$ s$^{-1}$.  By reviewing
other catalogs of clusters (David \etal\ 1993, hereafter D93; Ebeling
\etal\ 1998; DeGrandi \etal\ 1999; Jones \& Forman 1999) we have
augmented this list with five additional clusters that meet the flux
limit requirements: A1835, A2163, A3395, A3627 and Z5029. A3395 was
originally included in the sample (Edge \etal\ 1990), but was replaced
by A3391 in the revision (Edge, Stewart \& Fabian 1992). Our sample
includes both clusters.  Following the detailed treatment of the
completeness in Edge \etal\ (1990), we estimate that our augmented
sample is 80-100\% complete (60 of 60-75 clusters), with the Log N-Log
S and $\langle V/V_{max}\rangle$ distribution estimates giving
approximately 98\% completeness (60 of 61 clusters). We note that only
one of our additions to the sample (A3627) is located at low galactic
latitudes ($\rm|b|<20^\circ$).  For all but two of the sixty
clusters (A1644 and Z5029), pointed observations using the \asca\ GIS
were available, and it is these data which we then used to generate
our gas temperature maps.

\section{GENERATING THE TEMPERATURE MAPS\label{sec:methods}}
Spatially resolved spectroscopy is strongly dependent on the
Point-Spread-Function (PSF) in two ways. First is the loss of
sensitivity and angular resolution due to the scattering by the
telescope of photons from their nominal true locations.  This limits
the pixel to pixel contrast that can be reliably extracted from the
data. This is well known and understood for the photon intensity, but
applies equally well to spectral information.  The second effect
occurs when the PSF has a strong dependence upon energy, as was the
case with the \asca\ satellite (Takahashi \etal\ 1995). This leads to
a spatially dependent distortion of the derived spectra.  In the case
of a centrally concentrated extended source, such as a cluster of
galaxies, this can lead to spurious gradients in the derived
temperature and abundance distributions of the hot intracluster gas,
such that the outer regions will appear hotter with higher abundances
than actually present. The amplitude of this aliasing increases with
temperature and decreases with the angular size of the cluster. This
makes correction of the PSF of paramount importance for producing
reliable maps of the temperature distribution.

As a first step in the analysis, the \asca\ data were ``cleaned'' with
standard processing tools (Arnaud 1993).  A cutoff rigidity of 8
GeV/c, minimum Earth elevation angles of 5\d, and a maximum
count rate of 50 cts/s in the radiation belt monitor were used. The
GIS background was generated from an appropriately weighted
combination of background maps for all rigidities.

We then applied a technique developed by Churazov \etal\ (1996, 1999)
to correct the PSF and calculate the projected gas temperature
distribution. This method relies on the existence of a compact core in
the \asca\ PSF, and the assumption that the total PSF can be
represented as the sum of two components: a core PSF and an extended
``wings'' PSF.  The components are defined such that the wings contain
only a small fraction of the total photons detected and the core is
relatively small ($r<6^\prime$) in extent.  Then a direct correction
of the core PSF is combined with a Monte-Carlo simulation of the
effects of the wing component to produce the overall energy dependent
PSF.  The simplicity of this method makes its implementation
computationally fast, and it can be performed in each of the 1024
\asca\ GIS energy channels on a fine ($\approx 15''$) angular grid.

After correction for the PSF, the shape of the spectrum at a given
position is characterized as a linear combination of two template
spectra.  The two template spectra correspond to optically thin plasma
emission models at two different temperatures chosen so as to widely
bracket the nominal overall temperature of the cluster. These are
generated using a MEKAL model from XSPEC, convolved with the efficiency
of the telescope and the GIS detector. 

The two template spectra, although their heavy element abundances were
universally set to 0.3 solar, were individually tailored for each
cluster, by fixing the absorption column density to the appropriate
Galactic value and setting the redshift to the value found from
NED. We then attempted to generate a temperature map. If the template
spectra did not provide good bracketing for the cluster temperature,
we iterated until appropriate values for the templates were found.

The best fit weights of these template spectra needed to describe the
spectrum $S(E)$ observed in a given $15^{\prime\prime}$ pixel of the image are
then determined, i.e.
\begin{equation}
S(E)=A\cdot M(T_1,E)+B\cdot M(T_2,E), 
\end{equation} 
where $M(T,E)$ is the template spectrum for a given value of
temperature $T$, and the temperature is then calculated as a function
of the relative weights of the template spectra. Due to the linear
nature of the method, no complex fitting is required and, hence, the
method is computationally fast.  Although the limitations of this
method are significant (e.g., simple spectral forms must be assumed
{\em a priori}), this combined approach allows for reliable and rapid
generation of temperature maps specifically with the intent of
examining two dimensional spatial variations in gas temperature.

Churazov \etal\ (1996) have shown that an expectation value of the
temperature calculated this way is accurate (within a few percent of
the temperature obtained by conventional spectral fitting e.g. XSPEC)
under the assumption of a single temperature plasma with fixed
metallicity and absorption. This approach also has been previously
compared to other methodologies (Markevitch 1996a; Markevitch \etal\
1998, hereafter M98) and found to yield consistent results (A1367-
Donnelly \etal\ 1998; Coma- Donnelly \etal\ 1999).

Finally in generating the temperature maps, an adaptive smoothing has
been applied in order to reduce the noise throughout the image.  The
parameters of the adaptive smoothing have been adjusted so as to
produce comparable levels of uncertainty ($\sim 0.7$ keV at roughly
the 90\% confidence) across the entire range of temperatures and
intensities in our sample. As a last step, the intensity contours from
the \asca\ data are overlaid on the temperature maps.

\section{Results and Analysis\label{sec:results}}

Details of each observation as well as other ``global'' parameters for
each cluster are given in Table 1. The columns in the
table are as follows:
\begin{itemize}

\item {\em Column 1} gives the most common name for each cluster;
other common aliases for especially well known clusters (e.g. Coma)
are also listed here, while less common aliases are given in the notes
section.

\item {\em Columns 2 and 3} give the J2000 Right Ascension and
Declination as determined from centroiding the X-ray emission detected
by the \asca\ GIS.

\item {\em Column 4} is the redshift from the NED database.

\item {\em Columns 5 and 6} give the \asca\ sequence number(s) and
observation length(s).

\item {\em Column 7} gives the weighted value of the neutral hydrogen
column density within a 0.5\d\ cone centered on the cluster emission
from the FTOOL {\em nh}, in units of $10^{20}$ cm$^{-2}$.

\item {\em Column 8} is a global emission weighted temperature with a
90\% confidence error bar. This was found by fitting all of the \asca\
data within either a 1 Mpc or 20\dm -- whichever was smaller-- radius
aperture centered on the X-ray emission with a single Raymond-Smith
model. Point sources were not excluded, similar to the treatment in
M98. Because there were no \asca\ observations of A1644 and Z5029, the
result for A1644 is the {\em EINSTEIN} MPC result taken from D93 and
that for Z5029 is taken from Ebeling \etal\ (1998).

\item {\em Column 9} is the unabsorbed flux in the energy range from
2-10 keV within an aperture of 45\dm\ diameter in units of $10^{-11}\
ergs\ sec^{-1}\ cm^{-2}$.  The data, except for three clusters, were
measured by the {\em EINSTEIN} MPC and reported in D93 or Edge \etal\
(1992) catalogs.  For the the other three clusters (Z5029, A1835 and
A3627) estimates of their fluxes were derived from {\em ROSAT} PSPC
data and then converted to the MPC energy band (2-10 keV) using PIMMS.

\item {\em Column 10} is the 2-10 keV X-ray luminosity, in $10^{44}$
ergs sec$^{-1}$, found from applying the WMAP cosmology (Bennett
\etal\ 2003)--
$\Omega_m=0.27$, $\Omega_\Lambda=0.73$ and $H_0=71$ km sec$^{-1}$
Mpc$^{-1}$.

\item {\em Column 11} gives our structural index discussed below in
the text. If we found a variation in the gas temperature at the
$\ge$90\% confidence level, the first component (1,2,3) gives the
severity of the temperature change, while the second component (S,L)
gives the spatial extent of the variation.

\item {\em Column 12} lists the rank of this cluster by luminosity and
is to aid in locating the cluster in the collection of luminosity
ordered temperature maps (Figure 2).
\end{itemize}

%\placetable{tab:obsdat}

When a significant ($\ge$90\% confidence) variation in the gas
temperature was detected in a cluster, we characterized the variation
based upon severity and extent. Each of these two metrics was assessed
independently, and they are listed in the Structural Index column of
Table 1.  The first component (`1,2, or 3') gives the
severity of the temperature change: `1' is mild ( $2.4<\Delta T<4.3$
keV), `2' is strong ($4.2<\Delta T<5.8$ keV), and `3' is extreme
($\Delta T> 5.8$ keV).  The second component (`S,L') gives the spatial
extent of the temperature changes over the cluster, where `S'
indicates a small, localized variation (less than one quadrant of the
temperature map), while `L' notes a variation that extends over more
than one quadrant.

A comparison between our overall temperatures and those from D93 and
M98 is given in Figure 1.  We find that our
results are in good agreement with the 55 {\em EINSTEIN} MPC results
of D93.  The comparison to the thirty clusters from the M98 sample is
even better. For our three hottest clusters (A754, A2163 and
Triangulum Australis), we have plotted results from sources other than
M98. The temperature for A754 derives from a recent \chandra\ result
(Markevitch \etal\ 2003), that for Triangulum Australis from earlier
work (Markevitch 1996a) and for A2163 from Holzapfel \etal\
(1997). While four clusters (A401, A2029, A3571 and A3266, all
discussed in the Section~\ref{sec:notes}) lie off the 1:1 line in the
comparison with the M98 sample, the correlation between the two sets of
temperatures, derived from the same \asca\ data, but using very
different methodologies, is very tight.  We have also explored whether
there is any systematic effect due to our decreased aperture for
higher redshift ($z>0.042$) clusters and find no such effect.

%\placefigure{fig:temp_compare}

\subsection{The Temperature Maps\label{sec:tmaps}}
The maps of the gas temperature distribution for the entire flux
limited \asca\ sample are presented in Figure 2.  They
are ordered by decreasing X-ray luminosity (in the energy range from
2-10 keV) and all, except the very hot clusters A2163, A2319,
Ophiucus, and A754, have a common color scale for ease of
comparison. For these four the scale was shifted to accommodate their
peak temperatures. Although both the lower and upper limiting values
for the temperature range for these four clusters differ from the
rest of the sample, the stretch of the scale is the same for the
entire sample. To highlight that these four scales are different, the
location and orientation of the scale bar was changed to lie along the
vertical axis, rather than the horizontal axis, as for all the others.

The stretch of the scale was chosen such that one ``band'' of color is
approximately equivalent to the local uncertainty in the temperature
determination. This means that the presence of {\em four} colors
indicates a significant (greater than 90\% confidence) change in the
temperature.

%\placefigure{fig:tmaps}

\subsection{Notes for individual clusters\label{sec:notes}}
What follows are comments for individual clusters-- aliases,
significant changes in the global temperature, conversion of the
aperture radius to Mpc for nearby clusters ($z<0.042$) where the
global temperature was aperture limited, etc.  We have searched the
literature using the article links built into the \xmm\ Science
Archive and \chandra's WebChaSeR to compare our results to the most
recent, high resolution results.  Within our sample of sixty clusters,
we have found cluster gas temperature results based upon \chandra\
data for twenty-seven.  Six of these also have results published using
\xmm\ data, while a seventh (Coma) has \xmm\ results, but none from
\chandra. Although, the results are often difficult to compare, due to
the much higher angular resolution and the different fields of view,
especially for \chandra, we comment below upon the qualitative
comparisons between the \asca\ and \chandra /\xmm\ data sets. We note
that of this set of twenty-seven observed at higher resolution, only
seven (Perseus, A3266, A754, A1367, Virgo, Coma and A2256) cover
spatial extents comparable to our maps.

We also compare our results with previous work derived from \asca\
data but using different methodologies. Of the twenty-seven with
\chandra\ results, eleven had previous results (other than ours) from
\asca, while another twelve clusters with previous \asca\ measurements
have no published \chandra\ or \xmm\ analyses. Temperature maps for
the remaining eighteen clusters appear here for the first time.

\begin{itemize}
\item {\em A85:} Kempner, Sarazin \& Ricker (2002) have studied the
  6\dm$\times$6\dm\ region surrounding the ``southern sub-cluster''
  using \chandra\ data. Our temperatures and the sense of the
  gradients on the two sides of the subcluster are in good agreement
  with these results. Our results are also in good agreement with
  previous \asca\ results reported by M98. The \xmm\ results of Durret
  \etal\ (2003) deal with the extended filamentary emission outside
  our field of view.
\item {\em A119:} Our results are in good agreement with previous
  \asca\ results derived using a different methodology (M98).
\item {\em A262:} The global fit temperature was restricted to a
  20\dm\ (0.39 Mpc) radius. 
\item {\em AWM 7:} The global fit temperature was restricted to a
  20\dm\ (0.41 Mpc) radius. This cluster is also known as WBL 88. 
\item {\em A399:} Our results are in excellent agreement with previous
  \asca\ results (M98).
\item {\em A401:} Except in one region-- our results are in excellent
  agreement with previous \asca\ results (M98). A hot feature
  indicated in our map lies in region 9 from M98.
\item {\em A426/Perseus:} The global fit temperature was restricted to
  a 20\dm\ (0.43 Mpc) radius. This cluster is centered on NGC
  1275. Our results are consistent with both the high (inner 5\dm) and
  moderate (inner 9\dm) resolution maps of the gas temperature
  generated by Schmidt, Fabian \& Sanders (2002) with \chandra\
  data. Our results are also in excellent agreement with the \xmm\ map
  of the gas temperatures across the entire cluster (Churazov \etal\ 2003).
\item {\em 2A0335+96:} The global fit temperature was restricted to a
  20\dm\ (0.82 Mpc) radius. This cluster is also known as RXJ0338.6+0958.
\item {\em A478:} The azimuthally averaged radial temperature profile
  to $r=6$\dm\ from Sun \etal\ (2003) using \chandra\ data is
  consistent with the inner region of our map.
\item {\em A3266:} This cluster is also known as Sersic 40/06. Our
  map, previously reported in Henriksen, Donnelly \& Davis (2000), is
  consistent across the entire cluster with the \chandra\ data
  presented by Henriksen \& Tittley (2002). Our results are also in
  agreement with previous \asca\ results (M98). We note that although
  the 2D maps of the temperature distribution are in agreement with
  that of M98, the overall temperature given by M98 is significantly
  lower than ours ($7.7\pm 0.8$ keV versus $9.3\pm 0.4$ keV; see
  Figure 1).
\item {\em A496:} The global fit temperature was restricted to a
  20\dm\ (0.78 Mpc) radius. Our results are consistent both with the
  \chandra\ results (the inner 4\dm) from Dupke \& White (2003) as
  well as those from Tamura \etal\ (2001b) within a radius of 9\dm\
  from \xmm. Our results are also in agreement with previous \asca\
  results (Markevitch \etal\ 1999b). 
\item {\em 3C129.1:} The global fit temperature was restricted to a
  20\dm\ (0.53 Mpc) radius. Our results are consistent with the radial
  temperature profile (out to 19\dm) derived from the \chandra\
  data by Krawczynski (2002).
\item {\em A3391:} Our results are consistent with previous \asca\
  results (M98).
\item {\em A3395:} This cluster is also known as SC0627-54. Our
  results were previously reported in Donnelly \etal\ (2001). Our
  results are also in good agreement with previous \asca\ results
  (M98).
\item {\em A576:} The global fit temperature was restricted to a
  20\dm\ (0.91 Mpc) radius.
\item {\em PKS0745-19:} We note the large change in the fit global
  temperature from $8.5^{+1.9}_{-1.4}$ keV (D93) to
  $6.4^{+0.1}_{-0.2}$ keV.  Our results are consistent with the radial
  temperature profile and a 2D temperature map of the inner
  1.6\dm$\times$1.6\dm\ using \chandra\ data (Hicks \etal\ 2002).
\item {\em A644:} Our results are generally consistent out to 10\dm\
  with previously reported \asca\ data (Bauer \& Sarazin 2000). We
  note that their region 5 is cooler than what we find for this area,
  while their region 3 is hotter than our results.
\item {\em A754:} Our results are consistent across the entire cluster
  with the \chandra\ results from Markevitch \etal\ (2003). We also
  note that we are consistent with previous results from the same
  \asca\ data, but using a different methodology for correcting for
  the PSF and generating the temperature maps (Henriksen \& Markevitch
  1996). For the comparison to our overall temperature in
  Figure 1, we have used the \chandra\ result
  from Markevitch \etal\ 2003 ($10.0\pm 0.3$ keV) which is
  inconsistent at the 90\% confidence level with that found in M98
  ($9.0\pm 0.5$ keV).
\item {\em Hydra A:} Our results are in good agreement with \chandra\
  results for the inner 5\dm\ radius (Nulsen \etal\ 2002; David \etal\
  2001; McNamara \etal\ 2000).
\item {\em A1060:} The global fit temperature was restricted to a
  20\dm\ (0.31 Mpc) radius. This cluster is also known as Hydra. Our
  results are consistent with the high resolution \chandra\ map
  generated for the inner 1.5\dm\ by Yamasaki, Ohashi \& Furusho (2002).
\item {\em A1367:} The global fit temperature was restricted to a
  20\dm\ (0.53 Mpc) radius. Our results, previously reported in
  Donnelly \etal\ (1998), are in excellent agreement across the entire
  cluster with the \chandra\ temperature map (Sun \& Murray 2002b).
\item {\em Z5029:} This cluster is also known as ZwCl1215.1+0400.
\item {\em Virgo:} Due to the extreme proximity of this cluster (16
  Mpc) the data is restricted to only the core of the cluster emission
  around M87. The global fit temperature was restricted to a 20\dm\
  (0.088 Mpc) radius. Our results are in very good agreement with data
  from both \chandra\ and \xmm\ for the inner 16\dm$\times$16\dm\
  (DiMatteo \etal\ 2003; Molendi 2001; Gastadello \& Molendi 2002;
  B\"ohringer \etal\ 2001).
\item {\em A3526/Centaurus:} The global fit temperature was restricted to a
  20\dm\ (0.28 Mpc) radius. This cluster is also known as Klemola
  20. Our results are consistent with those found for the inner 4\dm\
  by Sanders \& Fabian (2002) with \chandra\ data. A similar map of
  the gas temperatures, developed with an earlier version of our
  software, appears in Churazov \etal\ (1999). 
\item {\em A3532:} This cluster is also known as Klemola 22.
  %forcomparisoncheckFlux=1.89e-11from RASS estimate.....
\item {\em A1651:} We find very good agreement with previous \asca\
  results (M98).
\item {\em A1656/Coma:} The global fit temperature was restricted to a
  20\dm\ (0.55 Mpc) radius. Our results, previously reported in
  Donnelly \etal\ (1999), are in good agreement with recent \xmm\
  results (Neumann \etal\ 2003) including the region of hot gas
  located to the north of the two central galaxies. We note that early
  \xmm\ results using just the EPIC/MOS detectors (Arnaud \etal\
  2001a), although in good general agreement elsewhere, did not find
  the hot region north of the central galaxies. At the same time, data
  using the EPIC/pn detectors (Briel \etal\ 2001) did find very small
  scale, hot features which they suggested might be the source of the
  `hot spot' that we find in the \asca\ data.
\item {\em A1689:} Our results are in good agreement with the
  \chandra\ radial temperature profile for the inner 3\dm\ (Xue \& Wu 2002). 
%\item{\emA1736:}notechangeintemperaturefrom4.6
\item {\em A3558:} This cluster is also known as Shapley 8. We note
  that the global temperature reported in D93 is significantly lower
  than our result ($3.8\pm 2.0$ keV versus $5.8^{+0.3}_{-0.2}$
  keV). Our results are excellent agreement with previous \asca\
  results (Markevitch \& Vikhlinin 1997a).
\item {\em A3571:} The global fit temperature was restricted to a
  20\dm\ (0.92 Mpc) radius. Our 2D results are in excellent agreement
  with those derived from \asca\ by M98. For the global temperature,
  our aperture is significantly larger (20\dm\ versus 16\dm) than that
  used to derive the M98 mean temperature. The offset in temperature
  (larger aperture, higher temperature) is similar to that found for
  A2029 (Figure 1).
\item {\em A1795:} Our results are consistent with both the \chandra\
  (Ettori \etal\ 2002; Markevitch, Vikhlinin \& Mazzotta 2001a) 
  and \xmm\ (Arnaud \etal\ 2001b; Tamura \etal\ 2001a) radial temperature
  profiles out to 4\dm\ and 12\dm\ respectively. 
\item {\em A1835:} Our results are in good agreement with the \xmm\
  radial profiles of the temperature to a radius of 7\dm\ (Majerowicz,
  Neumann \& Reiprich 2002; Peterson \etal\ 2001). At the largest
  radius ($r=3.3$\dm), the \chandra\ temperatures are higher than ours
  (10-16 keV versus 7-8.5 keV) (Schmidt, Allen \& Fabian 2001),
  although Markevitch (2002a) indicates that this disagreement is due
  to background flaring in the \chandra\ data.
\item {\em A2029:} Our results are consistent with the \chandra\
  azimuthally averaged radial temperature profile out to 4\dm\ (Lewis,
  Buote \& Stocke 2003; Lewis, Stocke \& Buote 2002). We are also in
  broad agreement with the low resolution azimuthal profile found with
  \asca\ data using a different methodology (Sarazin, Wise \&
  Markevitch 1998); we note that in our map the region of hot gas
  (Figure 2, panel 1, second row, far right) to the
  northwest coincides with their very hot annular wedge.  For the
  global temperature our aperture is significantly smaller (11\dm\
  versus 16\dm) than that used to derive the M98 temperature. The
  offset in temperature (larger aperture, higher temperature) is
  similar to that found for A3571 (Figure 1).
\item {\em A2052:} The global fit temperature was restricted to a
  20\dm\ (0.83 Mpc) radius. Our results are consistent with the
  azimuthally averaged radial profile from \chandra\ and, although we
  can not resolve the cool feature in the inner
  1.25\dm$\times$1.25\dm, we are in general agreement with the 2D
  distribution of temperatures within this region (Blanton \etal\
  2001; Blanton, Sarazin \& McNamara 2003).
\item {\em MKW 3s:} This cluster is also known as WBL 564.  Our
  results are in good agreement with the \chandra\ and \xmm\ radial
  profile ($r<10$\dm) and 2D map in the inner 4\dm$\times$4\dm\
  (Mazzotta \etal, 2002a).%+3.0keV?? 
\item {\em A2065:} The global temperature value reported by D93 is
  significantly higher than our derived value ($8.4^{+3.3}_{-1.8}$ keV
  versus $5.8\pm 0.2$ keV). Our results are in excellent agreement
  with previous \asca\ results (Markevitch, Sarazin \& Vikhlinin
  1999a).  %oldcentroid15:22:43,27:43.4
\item {\em A2063:} We note that the value reported by D93 is
  significantly higher than our result ($4.1^{+1.2}_{-0.8}$ keV versus
  $2.6^{+0.2}_{-0.1}$ keV). The global fit was restricted to a
  20\dm\ (0.83 Mpc) radius. 
\item {\em A2142:} Our results are in good agreement in the inner
  3\dm$\times$3\dm\ map from the \chandra\ data (Markevitch \etal\
  2000). Although we do not resolve the cold front features, we do
  find increasing temperatures at larger radii both to the west and
  east.
\item {\em A2147:} The global fit temperature was restricted to a
  20\dm\ (0.83 Mpc) radius. 
\item {\em A3627:} This cluster is also known as the Norma
  Cluster. The global fit temperature was restricted to a 20\dm\ (0.38
  Mpc) radius.
\item {\em A2163:} Our results are consistent with the 2D \chandra\
  temperature map of the inner 6\dm$\times$6\dm\ (Markevitch \&
  Vikhlinin 2001b), although we do not resolve the small cool features
  to the south and the detailed distribution of very hot gas to the
  east. We are also in agreement with previous \asca\ results
  (Markevitch 1996a).  We have included an overall temperature for
  this cluster ($11.2\pm 1.1$ keV from Holzapfel \etal\ 1997) in our
  comparisons with the results from M98
  (Figure 1). This result was derived from \asca\
  data but not included in the M98 data set.  %Thisindicatesthatthere
  %arenosystematiceffectsathighertemperaturesinourresults
  %comparedtoprevious\asca\work.  %positionchangefrom16:15:34-6:7.4
\item {\em A2199:} The global fit was restricted to a 20\dm\ (0.71
  Mpc) radius. Although we do not resolve the cold region found in the
  \chandra\ 2D map of the inner 3\dm$\times$3\dm\ (Johnstone \etal\
  2002), our results are consistent with the overall map as well as
  the azimuthally averaged radial temperature profile. Our results are
  consistent with those reported previously using \asca\ data
  (Markevitch \etal\ 1999b).
\item {\em Triangulum Australis:} Our result for the global
  temperature is significantly higher than that reported in D93
  ($10.7\pm 0.5$ keV versus $8.0\pm 1.4$ keV), and is consistent with
  the result from Markevitch \etal\ (1996b) of $10.3\pm 0.8$ keV. We
  note that this temperature is higher than the value ($9.5 \pm 0.7$
  keV) reported in M98.
\item {\em A2256:} Our results are in excellent agreement with the
  \chandra\ results spanning the entire cluster (16\dm$\times$16\dm)
  from Sun \etal\ (2002a). This includes the detection of hot gas to
  the south and southeast and cool gas to the west. Our results are
  also generally consistent with those previously reported in
  Markevitch (1996a) and Markevitch \& Vikhlinin (1997b) using \asca\
  data.
\item {\em Ophiucus:} The global fit temperature was restricted to a
  20\dm\ (0.67 Mpc) radius. Our \asca\ results are significantly
  higher than those reported in D93 ($12.1^{+0.6}_{-0.4}$ keV versus
  $9.0^{+0.8}_{-0.7}$ keV). Our results are in excellent agreement
  with previous results derived from \asca\ data using a different
  methodology (Watanabe \etal\ 2001).
%\item{\em A2255:}171231/645.6;temperaturechangefrom7.3
\item {\em 2319:} Our results are generally consistent with those
  reported previously using \asca\ data (Markevitch
  1996a).%19:20:4543:57.7
\item {\em Cygnus A:} Due to the extreme brightness of the AGN within
  the field of view, we determined the overall temperature by
  performing a two component fit (Raymond-Smith and Power Law) to the
  spectrum after excluding emission from the AGN to a radius of
  5\dm. This gives a temperature of $4.5\pm0.7$ keV which is
  consistent with the D93 value of $4.1^{+4.3}_{-1.3}$ keV. Comparison
  of our 2D map with other results is hampered for this one cluster
  due to the adaptive smoothing we have applied which increases the
  footprint of the bright point source. Away from the region
  surrounding the AGN, the agreement is very good with both \asca\ and
  \chandra\ (Markevitch \etal\ 1999a and Smith \etal\ 2002
  respectively) %oldVelocityfromOwensetal1997=0.0629
  %notepositionshiftfrom195844/4038.2
\item {\em A3667:} Although we do not resolve the sharp change in
  temperature , our results are in good agreement with the \chandra\
  results for the inner 16\dm$\times$16\dm\ (Mazzotta, Fusco-Femiano
  \& Vikhlinin 2002b; Vikhlinin, Markevitch \& Murray 2001). Our
  results are in excellent agreement with previous \asca\ results
  (Markevitch \etal\ 1999a).
\item {\em A2597:} Our results are in good agreement with the
  azimuthally averaged radial temperature profile out to 3\dm\
  generated by McNamara \etal\ (2001) using \chandra\ data. 
\item {\em A4038:} The global fit was restricted to a 20\dm\ (0.71 Mpc)
  radius. This cluster is also known as Klemola 44.
\item {\em A4059:} Our results are in agreement with the general
  findings of the hardness ratio map for the inner 6\dm$\times$6\dm\
  generated from \chandra\ data by Heinz \etal\ (2002). Our results
  are also in agreement with previous \asca\ results (M98).
\end{itemize}
Ten of our clusters (AWM7, Perseus, 3C129, PKS0745, A644, A3627,
Triangulum Australis, Ophiucus, A2319 and Cygnus A) lie at low
galactic latitudes ($\rm|b|<20^\circ$). It is possible that variations
in our temperature maps may be due to very localized, but strong,
variations in the Galactic hydrogen column density.  To test this
hypothesis, we examined in detail the derived values for the two low
latitude clusters that had the most extreme temperature variations:
A2319 and Ophiucus. We sampled the derived values of $n_H$ at the
maxima and minima of the temperature distributions. We find no
systematic variations which would account for the temperature
structures that we detect.
%thatthecolumndensityvariedbylessthan5\%
%aboveandbelowthemeanvalueforeachclusterandthattherewaslittle
%correlationwithtemperature.Forexample.inthecaseofA2319the
%$n_H$'sfortwohotregionstotheeastandsouthwestdifferbynearly
%10\%,whilethe$n_H$valueforthecoldregiontothewestisnearly
%identicaltothatofthesouthwesternhotregion.  
We note that because the resolution of the Galactic $n_H$ atlas' are
very low (typically of order a degree) that the average weighted
sampling available to us may not be sufficient to reveal structures in
the Galactic hydrogen that would produce the features demonstrated by
our temperature maps.

\section{DISCUSSION\label{sec:disc}}
One of the most striking features of the maps of the gas temperature
is the very high prevalence of structure. In our sample, 71\% (41 of
58 clusters) show significant variations in gas temperature at the
90\% confidence level. The leading model for large scale changes in
the intracluster gas temperature is the dynamical interaction
associated with two (or more) sub-cluster sized objects as they
collide and eventually merge. Jones \& Forman (1999) had previously
found, based on isophotal maps from {\em EINSTEIN}, that at least 40\%
of clusters showed features suggesting large scale dynamical
activity. Detecting structure from the X-ray surface brightness is
limited by considerations of the geometry of the subcluster-cluster
interaction. In contrast, gas temperature maps are not so constrained
and can provide a more complete inventory of cluster merging
phenomena.  As a consequence, there are several clusters (e.g. A1689,
A644 etc.), whose surface brightness distributions show no evidence
for ongoing merger activity, i.e. their intensity isophotes are
azimuthally symmetric. However, the temperature maps for these
clusters do show significant structure, indicating recent dynamical
activity.

We note that our estimates of the presence and severity of structure
in the temperature distribution are very conservative. This is
especially true for clusters with lower global temperatures. For
example a cool cluster ($\sim 2$ keV) with perturbed regions ($\sim 4$
keV) would not meet the significance criteria derived from our
adaptive smoothing for a `significant' variation in temperature, even
though the change ($\frac{\Delta T}{T}$) is 100\%. Further, the high
resolution results from \chandra\ and \xmm\ indicate that many
features in the gas are present on small angular scales. The \asca\
data, due to its much more limited angular resolution, cannot detect
these features. As a consequence the overall fraction of structure
that we find in our sample should be considered as a lower limit.

Another important aspect of our results is that the 2-dimensional
nature of our maps is more sensitive to temperature variations than
radial profiles. This is especially true for clusters with very
localized variations in temperature (`S' clusters e.g. A2142, A2029,
etc.) where fitting the temperature in annuli would tend to conceal
the temperature variations.

\subsection{Correlation of Temperature Structures and
  Luminosity\label{sec:tvl}} 
In Figure 3 we plot the percentage of each level of
temperature variation within the luminosity quintiles corresponding to
the panels in Figure 2. Each quintile has nearly the
same number of clusters (the second and third have eleven each, while
the others have twelve).
 
%\placefigure{fig:struc}

Figure 3 shows that as luminosity increases the
percentage of clusters with significant temperature variations also
increases. In fact the two most luminous quintiles have {\em no}
clusters that do not exhibit significant variation in gas
temperature. This has implications for cosmological studies of distant
clusters. Since most X-ray flux limited distant cluster samples are
necessarily dominated by luminous clusters, they are also very likely
to be dynamically unsettled. This will affect a variety of measures
including estimates of the mean temperatures and derived masses,
unless the clusters are observed for sufficiently long times with high
spatial resolution to allow the derivation of temperature maps.

Figure 3 also shows that the severity of the
fluctuations in temperature correlates with luminosity.  We find that
there are no class 3 clusters at low luminosities and no class 0
clusters at high luminosities.  From a qualitative point of view this
is not necessarily surprising. More luminous clusters are
intrinsically more massive and are also likely to be located at the
nodal center of several large scale structure filaments. Both of these
properties would increase the frequency and severity of mergers with
sub-cluster sized objects as well as the possibility of an equal mass
cluster-cluster event, similar to what we see in our sample.

In a similar vein we find that for the two most luminous quintiles
where all of the clusters have temperature variations, 75\% (9 of 12)
of the temperature variations are widespread (our `L'
classification). In the third and fourth quintiles, where seven of the
eleven clusters have temperature structures, the split is roughly
50-50 (III: four of seven, IV: three of seven are `L' class clusters )
between widespread (`L') and localized (`S') variations. For the
lowest luminosity quintile, where only three of the twelve clusters
have variations, two clusters have widespread variations and one is
localized. This suggests that the more massive clusters are more
likely to have merger activity which encompasses the entire cluster,
i.e. a merger of roughly equal sized masses, whereas less luminous
clusters absorb smaller subclusters.

\subsection{Correlation of Temperature Structures and Diffuse Radio Sources\label{sec:tvrh}}
Several authors (Giovannini, Tordi \& Feretti 2000; Kempner \& Sarazin
2001) have explored the connection between diffuse radio sources not
associated with specific galaxies (a.k.a. halos and relics) with
merger activity and the lack of a cooling flow structure in the
cluster. We have compared our sample with the halo/relic catalogs drawn
from the NVSS and WENSS surveys to explore correlations between our
structural index based on temperature maps and the presence of these
diffuse radio sources.

There are some limitations to the two radio catalogs that strongly
reduce the overlap with our cluster sample.  The NVSS sample is
constrained to objects with declinations above $\delta=-40^\circ$ and
due to baseline considerations is insensitive to structures at
redshifts less than $z=0.042$. The WENSS sample, although able to
detect sources as close as $z=0.01$, only covers the sky northward of
$\delta =30^\circ$. Applying these restrictions to our sample, there
are 31 clusters which are included in both our flux limited X-ray
sample and either the NVSS or the WENNS radio samples.  

Of these 31 clusters, 24 exhibit some level of significant temperature
structure. Only eight of the 31 (A85, A401, A754, A2142, A2163, A2256,
A2255 and A2319) have some form of a radio halo/relic, and all of
these exhibit temperature variations indicative of a merger.  Two
additional clusters from our X-ray sample (Coma and A3667, both with
temperature structure) also have either a halo or relic, but are
missed by the NVSS and WENSS surveys due to redshift and/or
declination considerations.  We have searched the literature for
detections of other radio halos or relics and found no others that are
also in our flux limited X-ray sample. This suggests that while the
presence of a radio halo/relic indicates temperature variations in the
gas, that the reverse is not necessarily true.

We then compared the presence of a radio halo/relic with our
structural index, specifically the eight most thermally perturbed
clusters (our class `2' and `3' clusters). While four clusters (A754,
A2163, A2256 and A2319) have either a radio halo or relic, the other
four clusters (A478, A2029, Cygnus A and A2244) do not. It is possible
a halo/relic exists in Cygnus A, but that the brightness of the AGN
located near the center of the cluster precludes detection of a
diffuse source. At the next lower level of temperature variation,
eleven class '1' clusters, which lie within the coverage of the NVSS
sample, do not have either a radio halo or relic detected within
them. However, the other six class '1' clusters (A85, A401, A2142 and
A2255, as well as Coma and A3667) do have a radio halo/relic.  This
suggests that the presence of a radio halo/relic does not correlate
with the severity of the variation in the gas temperature.

From this small sample, we find that one third of clusters with
temperature variations (eight of twenty four) have either a radio halo
or relic, and that the presence of a halo/relic does not correlate
with severity of the temperature structure.  Due to the suggested
connection between cluster merging and radio halos/relics (Kempner \&
Sarazin 2001), this highlights the potentially interesting exceptional
nature of those highly perturbed merging clusters that do not contain
a halo/relic source (i.e. A478, A2029, Cygnus A and A2244).

\section{Summary}
Maps of the gas temperature in galaxy clusters are a strong test for
the presence of dynamical effects, particularly mergers. This is
because they are not as dependent upon viewing geometry as the X-ray
surface brightness and the temperature variations are erased on longer
time scales than surface brightness variations (Schindler \& M\"uller
1993; Ricker 1998). The flux limited atlas of galaxy cluster
temperature maps presented here shows a rich diversity of structure
across the entire sample, particularly among the more luminous
clusters.  Many clusters previously thought of as ``old'' and relaxed
demonstrate significant variations in gas temperature, and there are
correlations with luminosity, and thus mass, both in the severity and
extent of the temperature variations.  This high frequency of
temperature variations, especially in luminous clusters at relatively
low redshift ($z< 0.2$), emphasizes the need for caution in utilizing
high redshift clusters to measure cosmological parameters. We find few
isothermal clusters in our flux limited sample. Thus either assuming
isothermality when deriving cluster masses or using the temperature as
a proxy for the mass in high redshift clusters can lead to erroneous
results (Markevitch \etal\ 2002b).

In contrast to the canonical model of little more than two decades ago,
merging activity in clusters appears to be the norm rather than the
exception. It is also clear from high resolution observations from
\chandra\ and \xmm\ that there is an exceptional amount of structure
that awaits detailed analysis and modeling.

\acknowledgments 

We thank Larry David for helpful conversations regarding the {\em
EINSTEIN} MPC data and A.C. Edge, J. Kempner, M. Markevitch and
P. Gorenstein for helpful suggestions.  We acknowledge support from
the Smithsonian Institute and NASA contract NAS8-39073.

\newpage

%\begin{deluxetable}{crrccccccccc}
\begin{longtable}{crrccccccccc}
\tabletypesize{\scriptsize}
\tablewidth{6.9in}
\tablecolumns{12}
\tablecaption{Observational Data}
\tablehead{Cluster&\multicolumn{1}{c}{$\alpha$\tablenotemark{a}}&\multicolumn{1}{c}{$\delta$\tablenotemark{a}}&$z$&Seq.
  No. &On Time&$n_H$\tablenotemark{b}&$kT$\tablenotemark{c}&$F$\tablenotemark{d}&$L$\tablenotemark{e}& Struc.&Index}
\startdata
A85            & 0:41:45& -9:19.6&0.0555&81024000&25111& 3.58& $6.7^{+0.1}_{-0.2}$&6.22& 3.59&1S&20\\%058
	       &        &        &      &81024010&10282&     &       		  &    &     &&\\
A119           & 0:56:15& -1:15.2&0.0442&83045000&33951& 3.10& $5.7^{+0.2}_{-0.2}$&2.75& 1.03&1L&42\\%053
A262           & 1:52:50& 36:08.8&0.0163&81031000&15861& 5.36& $2.1^{+0.1}_{-0.1}$&2.24& 0.12&--&59\\%035
AWM7           & 2:54:26& 41:34.8&0.0172&80036000&10374& 9.21& $3.8^{+0.2}_{-0.1}$&8.58& 0.52&--&54\\%036
A399           & 2:57:48& 13:03.2&0.0724&82008000&27942&10.60& $7.7^{+0.4}_{-0.4}$&3.42& 3.23&1L&22\\%032
A401           & 2:58:57& 13:34.3&0.0737&82010000&31264&10.30& $9.2^{+0.3}_{-0.4}$&5.73& 5.59&1L&11\\%038
	       &        &        &      &82009000&32374&     &        		  &    &     &&\\
A3112          & 3:17:58&-44:14.2&0.0750&81003000&36000& 2.55& $4.6^{+0.2}_{-0.1}$&1.94& 1.95&2L&31\\%104
A426/Perseus   & 3:19:49& 41:30.9&0.0179&80007000&12202&15.70& $5.1^{+0.1}_{-0.1}$&75.4& 4.94&2L&13\\%145
	       &        &        &      &80008000&19924&     &        		  &    &     &&\\
	       &        &        &      &83051000&12238&     &        		  &    &     &&\\
	       &        &        &      &83053000&13500&     &        		  &    &     &&\\
2A0335+96      & 3:38:39&  9:58.3&0.0349&82029000&17262&18.00& $3.1^{+0.1}_{-0.1}$&4.65& 1.11&2S&41\\%168
	       &        &        &      &82040000&36320&     &        		  &    &     &&\\
A3158          & 3:42:40&-53:37.8&0.0597&84020000&30101& 1.06& $5.6^{+0.2}_{-0.2}$&2.47& 1.63&2S&35\\%104
A478           & 4:13:26& 10:27.9&0.0881&81015000&32053&15.30& $6.4^{+0.3}_{-0.3}$&5.70& 7.68&2S&8\\%073
A3266          & 4:31:23&-61:25.0&0.0589&83023000&32299& 1.48& $9.3^{+0.4}_{-0.4}$&5.48& 3.53&2L&21\\%070
A496           & 4:33:39&-13:15.4&0.0329&80003000&32636& 4.56& $4.1^{+0.1}_{-0.1}$&5.30& 1.13&2S&39\\%112
3C129.1        & 4:49:59& 45:02.2&0.0223&86050000&38089&73.30& $5.9^{+0.1}_{-0.1}$&8.46& 0.85&1S&45\\%034
A3391          & 6:26:22&-53:41.3&0.0514&72019000&16945& 5.42& $5.7^{+0.4}_{-0.3}$&1.78& 0.89&1S&44\\%037
A3395          & 6:27:14&-54:28.5&0.0506&82033000&31065& 5.42& $4.7^{+0.2}_{-0.2}$&1.93& 0.94&1L&43\\%062
A576           & 7:21:31& 55:46.0&0.0389&84001000&41548& 5.69& $3.8^{+0.2}_{-0.1}$&1.97& 0.58&--&53\\%040
PKS0745-19     & 7:47:31&-19:17.6&0.1028&81016000&35834&43.80& $6.4^{+0.1}_{-0.2}$&5.71&10.14&1L&3\\%055
A644           & 8:17:23& -7:31.4&0.0704&83022000&54396& 6.76& $7.8^{+0.3}_{-0.3}$&4.13& 3.70&1S&18\\%040
A754           & 9:09:04& -9:39.9&0.0542&82057000&21598& 4.21&$11.8^{+0.8}_{-0.5}$&7.26& 4.00&3L&15\\%070
Hydra A        & 9:18:06&-12:05.9&0.0538&80015000&18714& 4.93& $3.7^{+0.2}_{-0.1}$&3.00& 1.63&--&36\\%045
A1060          &10:36:51&-27:31.6&0.0126&80004000&29701& 4.93& $3.1^{+0.1}_{-0.1}$&4.88& 0.16&--&58\\%045
A1367          &11:44:44& 19:43.0&0.0220&81029000&10624& 2.19& $3.7^{+0.1}_{-0.1}$&3.35& 0.33&2L&56\\%140
               &        &        &      &81029010& 8030&     &                    &    &     &&\\ 
               &        &        &      &81030000& 7292&     &                    &    &     &&\\
               &        &        &      &81030010& 9654&     &                    &    &     &&\\
Z5029          &12:17:41&  3:39.5&0.0750&    --- &  ---& 1.73& $6.3^e$        &$1.79^R$& 1.80&&33\\
Virgo*(core)   &12:30:48& 12:23.8&0.0036&60033000&11998& 2.64& $2.4^{+0.1}_{-0.1}$&30.0& 0.08&--&60\\%061
A3526/Centaurus&12:48:54&-41:18.3&0.0114&80032000&14710& 8.07& $3.5^{+0.1}_{-0.1}$&10.3& 0.28&1S&57\\%077
	       &        &        &      &80033000&11150&     &                    &    &     &&\\
	       &        &        &      &80034000&12368&     &                    &    &     &&\\
	       &        &        &      &83026000&57230&     &                    &    &     &&\\
A1644          &12:57:15&-17:21.2&0.0473& ---    & --- & 4.82& $4.7^{+0.9}_{-0.7}$&2.71& 1.16&&38\\
A3532          &12:57:20&-30:22.1&0.0554&86014000&23573& 5.96& $4.3^{+0.2}_{-0.2}$&1.95& 1.12&--&40\\%048
	       &        &        &      &86016000&24781&     &			  &    &     &&\\
A1650          &12:58:41& -1:45.5&0.0845&84021000&43394& 1.54& $5.8^{+0.2}_{-0.2}$&2.18& 2.73&1S&25\\%036
A1651          &12:59:23& -4:11.2&0.0844&82036000&35025& 1.71& $6.3^{+0.4}_{-0.3}$&3.67& 4.58&1L&14\\%037
A1656/Coma     &12:59:44& 27:56.3&0.0231&80016000&7462 & 0.89& $9.0^{+0.3}_{-0.4}$&25.1& 2.70&1L&26\\%036
A1689          &13:11:34& -1:21.9&0.1832&80005000&29324& 1.82& $9.4^{+0.7}_{-0.4}$&1.72& 8.13&1L&5\\%033
A1736          &13:26:59&-27:09.5&0.0458&83061000&16378& 5.36& $3.8^{+0.2}_{-0.1}$&1.74& 0.70&1L&49\\%042
A3558          &13:27:58&-31:30.2&0.0480&82046000&13967& 3.84& $5.8^{+0.3}_{-0.2}$&4.21& 1.85&--&32\\%029
A3562          &13:33:39&-31:40.5&0.0490&84041000&15365& 3.91& $5.0^{+0.3}_{-0.3}$&3.62& 1.65&--&34\\%021
A3571          &13:47:30&-32:51.7&0.0391&82047000&23886& 3.31& $8.3^{+0.3}_{-0.3}$&12.3& 3.66&1L&19\\%034
A1795          &13:48:49& 26:36.5&0.0631&80006000&31809& 1.18& $6.1^{+0.2}_{-0.1}$&5.19& 3.80&1S&17\\%041
A1835          &14:01:02&  2:52.7&0.2532&82052000&14932& 2.40& $7.7^{+0.5}_{-0.5}$&$1.75^R$&13.64&1L&1\\%043
               &        &        &      &82052010&13614&     &			  &    &    &&\\
A2029          &15:10:59&  5:44.8&0.0773&81023000&30409& 3.03& $7.1^{+0.3}_{-0.2}$&7.52& 8.00&2S&6\\%062
A2052          &15:16:46&  7:00.0&0.0350&85061000&35919& 2.78& $3.0^{+0.1}_{-0.1}$&2.48& 0.60&--&52\\%047
MKW3s          &15:21:50&  7:42.4&0.0450&80011000&24806& 3.01& $3.4^{+0.2}_{-0.1}$&1.71& 0.66&--&50\\%049
A2065          &15:22:27& 27:42.8&0.0726&84054000&23430& 2.87& $5.8^{+0.2}_{-0.2}$&2.75& 2.61&1L&28\\%047
	       &        &        &      &84054010&20382&     &                    &    &     &&\\
A2063          &15:23:07&  8:36.8&0.0353&81002000&20000& 2.92& $2.6^{+0.2}_{-0.1}$&2.46& 0.60&--&51\\%061
A2142          &15:58:23& 27:14.3&0.0909&81004000&13878& 4.05& $8.7^{+0.6}_{-0.6}$&6.77& 9.65&1S&4\\%031
A2147          &16:02:19& 15:59.0&0.0350&83074000&30613& 3.46& $4.5^{+0.1}_{-0.2}$&2.96& 0.71&--&48\\%050
A3627          &16:14:08&-60:52.2&0.0157&84005000&35394&20.70& $5.4^{+0.1}_{-0.1}$&$15.7^R$& 0.80&2L&47\\%083
               &        &        &      &84005010&24269&     &                    &    &     &&\\ 
A2163          &16:15:45& -6:08.2&0.2030&80024000&26628&12.30&$11.5^{+0.9}_{-0.8}$&2.33&12.97&2L&2\\%046
A2199          &16:28:38& 39:33.3&0.0299&80023000&27433& 0.87& $4.4^{+0.1}_{-0.1}$&6.94& 1.23&--&37\\%046
A2204          &16:32:46&  5:34.6&0.1523&82045000&13728& 5.94& $7.1^{+0.2}_{-0.3}$&2.20& 7.68&1L&7\\%053
	       &        &        &      &82045010&15294&     &                    &    &     &&\\
Tri. Aust.     &16:38:16&-64:21.2&0.0510&83060000&11554&12.30&$10.7^{+0.5}_{-0.5}$&11.0& 5.41&1L&12\\%041
               &        &        &      &83060010& 6762&     &                    &    &     &&\\
A2244          &17:02:39& 34:04.0&0.0968&86073000&29111& 2.07& $5.6^{+0.2}_{-0.2}$&1.68& 2.68&3S&27\\%110
A2256          &17:04:01& 78:37.9&0.0581&10004020&8764 & 4.11& $7.4^{+0.2}_{-0.1}$&5.05& 3.17&2L&23\\%070
	       &        &        &      &10004030&23652&     &                    &    &     &&\\
	       &        &        &      &80002000&33908&     &                    &    &     &&\\
Ophiuchus      &17:12:24&-23:21.0&0.0280&80027000& 7262&21.60&$12.1^{+0.6}_{-0.4}$&43.7& 6.84&3L&10\\%053
A2255          &17:12:50& 64:03.7&0.0806&84012000&39173& 2.55& $6.3^{+0.3}_{-0.2}$&1.71& 1.96&1L&30\\%049
	       &        &        &      &84012010&31333&     &                    &    &     &&\\
A2319          &19:21:09& 43:59.3&0.0557&80041000&13458& 7.02& $9.9^{+0.4}_{-0.4}$&12.1& 7.02&3L&9\\%064
	       &        &        &      &80041010&11662&     &                    &    &     &&\\
Cygnus A       &19:59:24& 40:45.6&0.0569&70003000&22006&31.30& $6.0^{+0.1}_{-0.4}$&4.78& 2.78&2L&24\\%117
	       &        &        &      &70003010&28660&     &                    &    &     &&\\
A3667          &20:12:25&-56:49.6&0.0556&83054000&17920& 4.77& $7.1^{+0.3}_{-0.3}$&6.68& 3.86&1L&16\\%034
A2597          &23:25:17&-12:07.6&0.0852&83062000&40230& 2.46& $3.7^{+0.1}_{-0.1}$&1.90& 2.41&--&29\\%024
A4038          &23:47:41&-28:08.7&0.0300&83004000&45581& 1.55& $3.1^{+0.1}_{-0.1}$&2.55& 0.46&--&55\\%061
A4059          &23:57:03&-34:45.1&0.0475&82030000&33385& 1.10& $4.2^{+0.2}_{-0.1}$&1.88& 0.81&--&46\\%036
\enddata
\tablenotetext{a}{Coordinates are J2000.}
\tablenotetext{b}{ $\times 10^{20}$ cm$^{-2}$. Weighted Galactic n$_H$ within
  0.5\d\ radius cone from the FTOOL {\em nh}.}      
\tablenotetext{c}{ keV. Single component Raymond-Smith model of
  composite spectrum within a radial distance of 1 Mpc or 20\dm,
  whichever was smaller. The temperature for A1644 is the result from
  D93 using MPC data; for Z5029 the result is an estimate
  based on the {\em ROSAT} PSPC flux and the $L_X$-kT relation
  (Ebeling \etal\ 1998).}
\tablenotetext{d}{ $\times 10^{-11}$ ergs sec$^{-1}$ cm$^{-2}$. Unabsorbed
  flux within a radial distance of $22.5^\prime$ from the {\em
  EINSTEIN} MPC (D93). For clusters noted with a $^R$ the
  estimates are from the {\em ROSAT} PSPC and PIMMS.}
\tablenotetext{e}{ $\times 10^{44}$ ergs sec$^{-1}$. Luminosities found
  using the tabulated fluxes and distances assuming a {\em WMAP} cosmology.}
\end{longtable}

\end{document}